\documentclass[floats,prd,aps,showpacs,amssymb,nofootinbib,onecolumn]{revtex4}
\usepackage{amssymb}
\usepackage{graphics}
\usepackage{amsmath}
\usepackage{amsfonts}
\usepackage{bm}
\usepackage[]{latexsym}
\usepackage{mathtools}
\usepackage{xcolor}
\newcommand{\mb}{\mathbf}
\newcommand{\be}{\begin{equation}}\newcommand{\ee}{\end{equation}}
\newcommand{\bea}{\begin{eqnarray}}\newcommand{\eea}{\end{eqnarray}}
\newcommand{\brr}{\begin{array}}\newcommand{\err}{\end{array}}
\newcommand{\bit}{\begin{itemize}}\newcommand{\eit}{\end{itemize}}
\newcommand{\ben}{\begin{enumerate}}\newcommand{\een}{\end{enumerate}}

\newcommand{\ba}{\begin{array}}

\newcommand{\ea}{\end{array}}

\def\al{\alpha}

\def\1{{_{1}}}\def\2{{_{2}}}

\def\noHe0{:\;\!\!\;\!\!:H_e(0):\;\!\!\;\!\!:}
\def\noHm0{:\;\!\!\;\!\!:H_\mu(0):\;\!\!\;\!\!:}
\def\al{\alpha}

\def\1{{_{1}}}\def\2{{_{2}}}

\begin{document}
\title{Heuristic derivation of Casimir effect in minimal length theories}

\author{M.~Blasone\footnote{blasone@sa.infn.it}$^{\hspace{0.3mm}1,2}$, G.~Lambiase\footnote{lambiase@sa.infn.it}$^{\hspace{0.3mm}1,2}$,
G.~G.~Luciano\footnote{gluciano@sa.infn.it}$^{\hspace{0.3mm}1,2}$, L.~Petruzziello\footnote{lpetruzziello@na.infn.it}$^{\hspace{0.3mm}1,2}$ and F.~Scardigli\footnote{fabio@phys.ntu.edu.tw}$^{\hspace{0.3mm}3,4}$}

\vspace{2mm} \affiliation
{$^1$Dipartimento di Fisica, Universit\`a di Salerno, Via Giovanni Paolo II, 132 I-84084 Fisciano (SA), Italy.\\ $^2$INFN, Sezione di Napoli, Gruppo collegato di Salerno, Italy.\\
$^3$Dipartimento di Matematica, Politecnico di Milano,Piazza Leonardo da Vinci 32, 20133 Milano, Italy.\\ $^4$Institute-Lorentz for Theoretical Physics, Leiden University, P.O. Box 9506, Leiden, The Netherlands}

\date{\today}
\def\be{\begin{equation}}
\def\ee{\end{equation}}
\def\al{\alpha}
\def\bea{\begin{eqnarray}}
\def\eea{\end{eqnarray}}

\begin{abstract}
We propose a heuristic derivation of 
Casimir effect in the context of minimal length theories 
based on a Generalized Uncertainty Principle (GUP). By considering 
a GUP with only a quadratic term in the momentum, 
we compute corrections to the standard formula of Casimir energy
for the parallel-plate geometry, the sphere and the cylindrical shell. 
For the first configuration, we show that 
our result is consistent with the one
obtained via more rigorous calculations in 
Quantum Field Theory. Experimental developments 
are finally discussed.
\end{abstract}

\vskip -1.0 truecm 

\maketitle


\section{Introduction}	
Quantum vacuum is not empty. 
Seen up close, it is crowded with 
all sort of virtual particles continuously 
popping in and out of existence. 
One of the most outstanding manifestations of 
vacuum fluctuations is the Casimir effect~\cite{Casimir,Miltonbook},
which has recently aroused great interest in a large class of 
domains, ranging from quantum computing~\cite{Benenti} 
to biology~\cite{Andersen}.
As well known, the Casimir effect originates from alterations
of the zero-point energy induced by boundary conditions. 
The ensuing attractive
force is obtained by differentiating 
the vacuum energy with respect to the separation  
between the boundaries. In passing, we mention that 
an alternative derivation was  
proposed in Ref.~\cite{Jaffe}, where the Casimir effect
was addressed by considering relativistic 
van der Waals forces 
between metallic plates.

Besides its intrinsic interest in the standard
Quantum Field Theory (QFT), the Casimir 
effect provides a useful test bench for 
physics beyond the Standard Model~\cite{Blasone:2018obn} and gravity theories~\cite{Sorge,Petruz,Buoninf}.
In Refs.~\cite{Blasone:2018obn}, for instance,  
it was analyzed in connection
with the unitary inequivalence between mass and
flavor Fock spaces for mixed fields. 
Similarly, in Refs.~\cite{Petruz} and~\cite{Buoninf} the computation
of the Casimir energy density and pressure
was exploited to fix some constraints
on the characteristic free parameters 
appearing in the Standard Model Extension and in 
extended theories of gravity, respectively. 
Recently, extensive studies were also carried out 
in the context of the Generalized Uncertainty
Principle (GUP)~\cite{Nouicer,Harbach,Panella,Panella2,Dorsch}, where 
non-trivial corrections 
were shown to arise due to the existence
of a minimal length at Planck scale.
The present contribution fits in the last of the above
lines of research, since it aims to investigate
the connection between the Casimir effect and 
models which inherently embed this fundamental scale.

The concept of minimal length naturally 
emerges in quantum gravity theories in the form of an effective
minimal uncertainty in position $\Delta x_{\mathrm{min}}>0$. 
Several different theoretical arguments (many of them in form of Gedanken experiments) show the 
impossibility to measure arbitrarily short distances, due to the very existence of gravity. This naturally leads to a modification of the Heisenberg position-momentum uncertainty principle (HUP). In a
one-dimensional setting, studies of string theory, loop quantum gravity, 
deformed special relativity and black hole physics~\cite{VenezGrossMende,MM, MM1bis,MM2,FS,Adler2,CGS,SC2013}
have converged on the idea that a
proper generalization of the HUP would be
\begin{equation}
\Delta x\, \Delta p
\geq
\frac{\hslash}{2}
\left[1
+\beta
\left(\frac{\Delta p\, c}{E_p}\right)^2
\right],
\label{gup}
\end{equation}
where $E_p$ is the Planck energy and
we are retaining only the
leading-order correction in the 
dimensionless parameter $\beta>0$. Of course, in the limit $\beta\rightarrow0$,
the HUP of ordinary quantum mechanics is recovered, 
as it should be. 
Let us also remark that the deformation parameter $\beta$ is not
fixed by the theory: in principle, 
it can be either constrained via experiments~\cite{Brau:1999uv} 
or estimated by computational techniques in different contexts~\cite{Theor}, which yield $\beta\sim\mathcal{O}(1)$ (for a recent overview
on the various attempts to fix $\beta$, see Ref.~\cite{ScardRev}).  
However, given the high-energy scale
at which modifications of the HUP should become relevant, the natural arenas 
for testing  GUP effects are undoubtedly Hawking~\cite{ONG} and Unruh~\cite{SBLC} radiations.

Note that, for mirror-symmetric states 
(with $\langle \hat{p} \rangle = 0$), 
Eq.~\eqref{gup} can be equivalently 
rephrased in terms of the generalized commutation relation
\begin{equation}
\left[\hat{x},\hat{p}\right]
\,=\,
i\hslash \left[
1
+\beta
\left(\frac{\hat{p}\,c}{E_p} \right)^2 \right],
\label{gupcomm}
\end{equation}
since $\Delta x\, \Delta p \geq (1/2)\left|\langle [\hat{x},\hat{p}] \rangle\right|$.
Vice-versa, the above relation implies the inequality~(\ref{gup}) for any state. 
Moreover, in $n$ spatial dimensions, the commutator~\eqref{gupcomm}
can be cast in different forms, among which the most common is 
\begin{equation}
\label{moregeneralform}
\left[\hat{x}_i,\hat{p}_j\right]
=\left[f(\hat p^2)\,\delta_{ij}\,+\,g(\hat p^2)\,\hat p_i\,\hat p_j\right],\,\quad i,j=1,\dots,n
\end{equation}
with $f(\hat p^2)=1+\beta
\left(\frac{\hat{p}\,c}{E_p} \right)^2$ and $g(\hat p^2)=0$ (note that in 
$n$-dimensions, the functions $f(\hat p^2)$ and $g(\hat p^2)$ can be
chosen in different ways~\cite{Panella,Panella2}. In any case, they are not completely arbitrary, 
being related via the requirement of translational and/or rotational symmetry of the
commutator~\cite{Sc1}).

Working in the outlined scenario, in the present paper we
compute GUP-corrections 
to the Casimir energy for three different geometries:
the parallel-plate configuration, the spherical and cylindrical shells.
For the first case, we follow a
field theoretical treatment first~\cite{Nouicer,Harbach,Panella,Panella2} and then
a heuristic derivation. The two approaches  
are found to be consistent as concerns the dependence
of the corrective term on the inverse fifth power
of the distance between the plates.
On the other hand, to the best of our knowledge,  
this is the first time that the Casimir effect for spherical and cylindrical
geometries is addressed in the context of GUP. Thereby, we can only
compare our results with the ones existing in literature 
in the limit of vanishing $\beta$.

The remainder of the work is organized as follows: in Section~\ref{CaQFT}
we briefly review the standard calculation of the Casimir
vacuum energy for the parallel-plate geometry.
The obtained result is then extended to the
context of GUP by quantizing the field in
the formalism of maximally localized states.
Motivated by the utility of heuristic procedures which help to develop physical
intuition, in Section~\ref{CaHe} we 
consider a similar derivation of the Casimir
effect both from HUP and GUP based on simple quantum 
and thermodynamic arguments. In this regard, we also clarify
the approach of Ref.~\cite{Gine}, where the Casimir effect
is deduced from the HUP by naively introducing
an effective radius $r_e$. The above reasoning is then 
applied to the cases of a spherical and cylindrical shells in Sec.~\ref{sphcylshel}. 
Finally, 
conclusions and perspectives  
are given in Section~\ref{DandC}.

\section{Casimir effect for parallel plates: QFT approach}
\label{CaQFT}
In the framework of canonical QFT, the Casimir
effect can be derived via different approaches and
in a wide range of contexts. By referring 
to the original treatment by Casimir, in what follows
we sketch the main steps leading 
to the relation between the zero-point energy $\Delta E$
and the distance $d$ between the plates. 

\subsection{Casimir effect from Heisenberg uncertainty principle}
We consider the simplest three-dimensional 
geometry of two parallel plates separated 
by a distance $d$ along the $x$-axis. Let $L$ be the side of  
the plates (with $L\gg d$) and $S=L^2$ their surface area. The Casimir
effect arises from the vacuum fluctuations
of any quantum field in the presence
of such boundary conditions on field modes. Consider, for example, the 
electromagnetic field 
$\hat{\mathbf{A}}\hspace{0.2mm}(t,\mathbf{x})$ in the 
Coulomb gauge $\mathbf{\nabla}\cdot \mathbf{A}=0$,
\begin{equation}
\label{elefield}
\hat{\mathbf{A}}(t,\mathbf{x})\,=\,\sum_{\lambda=1,2}\int\frac{d^3p}{{(2\pi)}^3}\sqrt{\frac{{(2\pi)}^4\,\hslash c^2}{\omega_{p}}}\left[\epsilon_{\mathbf{p},\lambda}\,\hat a_{\mathbf{p},\lambda}\,\psi_{\mathbf{p}}(t,\mathbf{x})\,+\,\mathrm{h.c.}\right],
\end{equation}
where $\epsilon_{\mathbf{p},\lambda}$ are the 
polarization vectors satisfying the relation $\epsilon_{\mathbf{p},\lambda}\,\epsilon^*_{\mathbf{p},\lambda'}=\delta_{\lambda\lambda'}$ and
$\psi_{\mathbf{p}}(t,\mathbf{x})$ 
are the plane waves (i.e. the standard position representation of momentum eigenstates) of frequency $\omega_{p}=cp/\hslash$.
The ladder operators $\hat a_{\mathbf{p},\lambda}$ in Eq.~\eqref{elefield} obey the canonical 
commutation relations.

Now, the vacuum energy responsible
for the attractive force between the plates
can be obtained by subtracting the infinite vacuum
energy of the electromagnetic field in free space
from the corresponding infinite energy between the 
perfectly conducting boundaries. Mathematically speaking, we have
\begin{equation}
\label{Casimirenergy}
\Delta E(d)\,\equiv\,E(d)\,-\,E_0\,=\,\langle0|\hat{H}(d)\,-\,\hat{H}|0\rangle\,,
\end{equation}
where the Hamiltonian is $\hat H=\frac{1}{8\pi}\int d^3x\,\big[{\big(\partial_0\hat{\mathbf{A}}\big)}^2-\hat{\mb{A}}\cdot\mathbf{\nabla}^2\hat{\mb{A}}\big]$.
By using this relation, one can show that~\cite{Panella} 
\begin{equation}
\label{eps}
\Delta E(d)=c\hspace{0.3mm}S\hspace{-0.8mm}\int \frac{d^2p_{\perp}}{{(2\pi\hslash)}^2}\left[\frac{|\mb{p}_\perp|}{2}+\sum_{n=1}^{\infty}\sqrt{{|\mb{p}_\perp|}^2+\frac{n^2\pi^2\hslash^2}{d^2}}-\int_{0}^{\infty}dn\,\sqrt{{|\mb{p}_\perp|}^2+\frac{n^2\pi^2\hslash^2}{d^2}}\right]\hspace{-0.2mm},
\end{equation}
where $\mb{p}_{\perp}=(p_y,p_z)$ is the transverse momentum
and we have exploited the fact that the
conditions of vanishing field on the plates only allow
for a discrete set of values of the
momentum along the $x$-axis, i.e. 
$p_x=\frac{n\pi\hslash}{d},$ with $n$ integer.
Note that, in the above calculations, we have 
neglected surface corrections. 
The integral in Eq.~\eqref{eps} is divergent for large values of 
the momentum. A possible trick to remove this infinity is to
introduce an ultraviolet momentum cutoff $p_{\mathrm{max}}\sim \hslash/d$
and remove the regularization only at the end of
calculations\footnote{Note that other common
regularization techniques are the zeta function regularization
and the point splitting~\cite{Bordag}.}. By following this procedure
and applying the asymptotic Euler--MacLaurin
summation formula, we obtain the well-known 
expression for the energy shift
\begin{equation}
\label{Casenedens}
\Delta E(d)\,=\,-\frac{\pi^2}{720}\frac{\hslash\hspace{0.2mm}c\hspace{0.2mm}S}{d^3}\,.
\end{equation}
For later convenience, we also write down the
formula for the Casimir energy in one spatial dimension
\begin{equation}
\label{casundim}
\Delta E(d)\,=\,-\frac{\pi}{12}\frac{\hslash\hspace{0.2mm}c}{d}\,.
\end{equation}

\subsection{Casimir effect in minimal length QED}
Let us now investigate the Casimir effect
in the presence of the generalized commutator~\eqref{moregeneralform}. 
Note that, in this case, a minimization of 
the generalized uncertainty 
relation with respect to $\Delta p_i$
gives the following nonzero minimal length $(\Delta x_i)_{\mathrm{min}}=\sqrt{\beta}\hspace{0.2mm}\ell_p$, where $\ell_p=\hslash c/E_p$ denotes the
Planck length.

A quantum theoretical framework which implements
the appearance of a nonzero minimal uncertainty
in position has been described in Ref.~\cite{MM1bis}.
Unlike the standard quantum mechanics,  
in this context we do not have
localized functions in the $\mathbf{x}$-space, 
so we have to introduce the so-called quasi-position
representation. This consists in projecting 
the state of the system onto the set of
\emph{maximally localized states}, which, by definition,
are characterized by the minimal position uncertainty $(\Delta x)_{\mathrm{min}}$.
In the momentum representation, the general (i.e. time-dependent) maximally
localized state around the average
position $\textbf x$ takes the form
\begin{equation}
\label{maxlocstat}
\widetilde{\psi}_{\mathbf{p}}(t,\mathbf x)\,=\,\frac{1}{{(\sqrt{2\pi\hslash})}^3}\,e^{-i\left[\widetilde\omega_p\,t\,-\,\widetilde{\mb{p}}\cdot\mb{x}/\hslash\right]},
\end{equation}
where 
\begin{equation}
\label{defpom}
\widetilde\omega_p\,=\,\frac{E_p}{\hslash\sqrt{\beta}}\arctan\left(\frac{cp\sqrt{\beta}}{E_p}\right)\hspace{-0.3mm},\quad\widetilde{\mathbf p}_i\,=\,\left[\frac{E_p}{cp\sqrt{\beta}}\arctan\left(\frac{cp\sqrt{\beta}}{E_p}\right)\right]\mathbf p_i\,.
\end{equation}
Note that, for $\beta\rightarrow 0$, 
the quasi-position representation reduces to the standard 
plane-wave formalism, since 
$\widetilde\omega_p\rightarrow\omega_p$, $\widetilde{\mathbf p}_i\rightarrow \mathbf{p}_i$ and
$(\Delta x_i)_{\mathrm{min}}=0$. 

In terms of the maximally localized states, the electromagnetic
field reads
\begin{equation}
\label{newfieldexp}
\hat{\mathbf{A}}(t,\mathbf{x})\,=\,\sum_{\lambda=1,2}\int\frac{d^3p}{{(2\pi)}^3\left(1+\frac{c^2p^2\beta}{E^2_p}\right)}\sqrt{\frac{{(2\pi)}^4\,\hslash^2 c^2\sqrt{\beta}}{E_p\arctan\left(\frac{cp\sqrt{\beta}}{E_p}\right)}}\left[\epsilon_{\mathbf{p},\lambda}\,\hat a_{\mathbf{p},\lambda}\,\widetilde\psi_{\mathbf{p}}(t,\mathbf{x})\,+\,\mathrm{h.c.}\right],
\end{equation}
where the factor $\left(1+c^2p^2\beta/E_p^2\right)^{-1}$
arises from the modified completeness relation for the momentum
eigenstates $|\mathbf{p}\rangle$.

In order to derive GUP-corrections to the Casimir effect, 
let us replace the field expansion~\eqref{newfieldexp}
in the Hamiltonian. 
The calculation of the energy shift~\eqref{eps} 
proceeds as usual~\cite{Panella2}. 
The remarkable difference, however, is that in this case we
do not need to introduce
any restriction on the momentum scale. 
Indeed, from Eq.~\eqref{defpom}, it follows the natural cutoff
$\widetilde{p}_{\mathrm{max}}=\pi E_p/(2c\sqrt{\beta})$, 
leading to $n_{\mathrm{max}}=E_pd/(2\hslash c\sqrt{\beta})$.
Accordingly, the Casimir energy takes the form~\cite{Panella2}
\begin{equation}
\label{veinfs}
\Delta E(d)\,=\,-\frac{\pi^2}{720}\frac{\hslash\hspace{0.2mm}c\hspace{0.2mm}S}{d^3}\left[1\,+\,\frac{2\pi^2\hspace{0.2mm}\beta}{3}{\left(\frac{\hslash c}{E_pd}\right)}^2\right],
\end{equation}
to be compared with the standard QED expression~\eqref{Casenedens}.
Note that the GUP term is attractive, since it
contributes to increase the modulus of the energy. 

\begin{figure}[t]
 \centering
 \includegraphics[width=12cm]{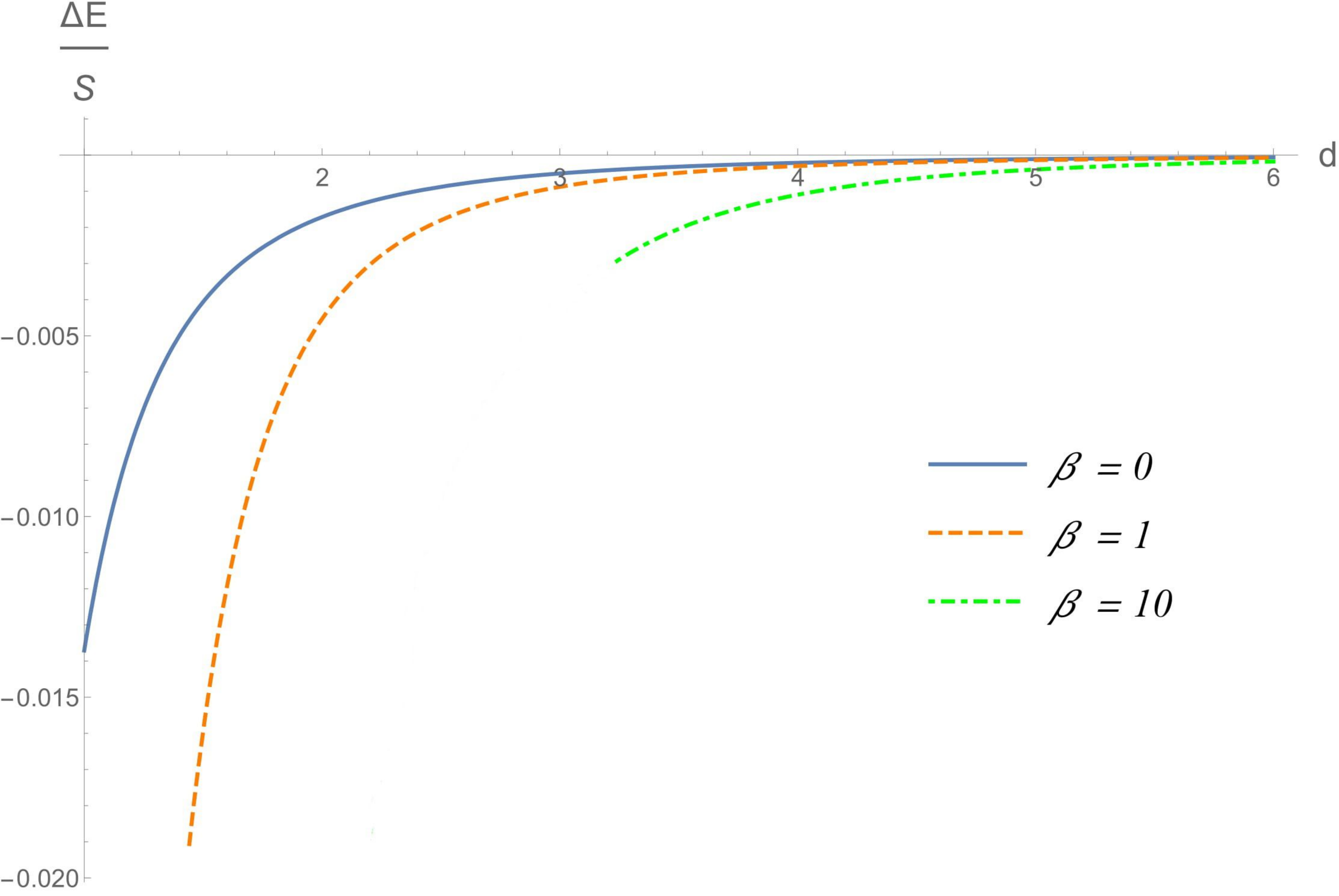}
 \renewcommand{\baselinestretch}{0.8}
 \caption{Casimir energy per unit surface versus the distance between the plates for different values of $\beta$ (quantities are in Planck units). Note that the green (dash-dotted) line starts from the minimal distance $d=\sqrt{\beta}\hspace{0.2mm}\ell_p\simeq3\hspace{0.3mm} \ell_p$.}
 \label{figura1}
 \end{figure}

In Fig.~\ref{figura1}, the 
Casimir energy per unit surface area has been plotted as a function of the distance
between the plates for three values of the deformation parameter $\beta$. For distances
large enough, the different curves overlap, being the effects
of the minimal length negligible.

\section{Casimir effect for parallel plates: heuristic approach}
\label{CaHe}
Although the heuristic arguments we 
are going to discuss do not give the exact
expression for the Casimir force, 
they allow us to better understand 
the origin of this effect, as well as
the nature of GUP-corrections to the standard formula~\eqref{Casenedens}. 
Thus, following the guidelines
of the previous Section, we provide a heuristic computation 
of Casimir energy by using the Heisenberg 
Uncertainty Principle first and then working in the framework of minimal length 
theories based on the Generalized Uncertainty Principle~\eqref{gup}.
Comparison with the corresponding QFT results is finally discussed.

\begin{figure}[t]
 \centering
 \includegraphics[width=7.5cm]{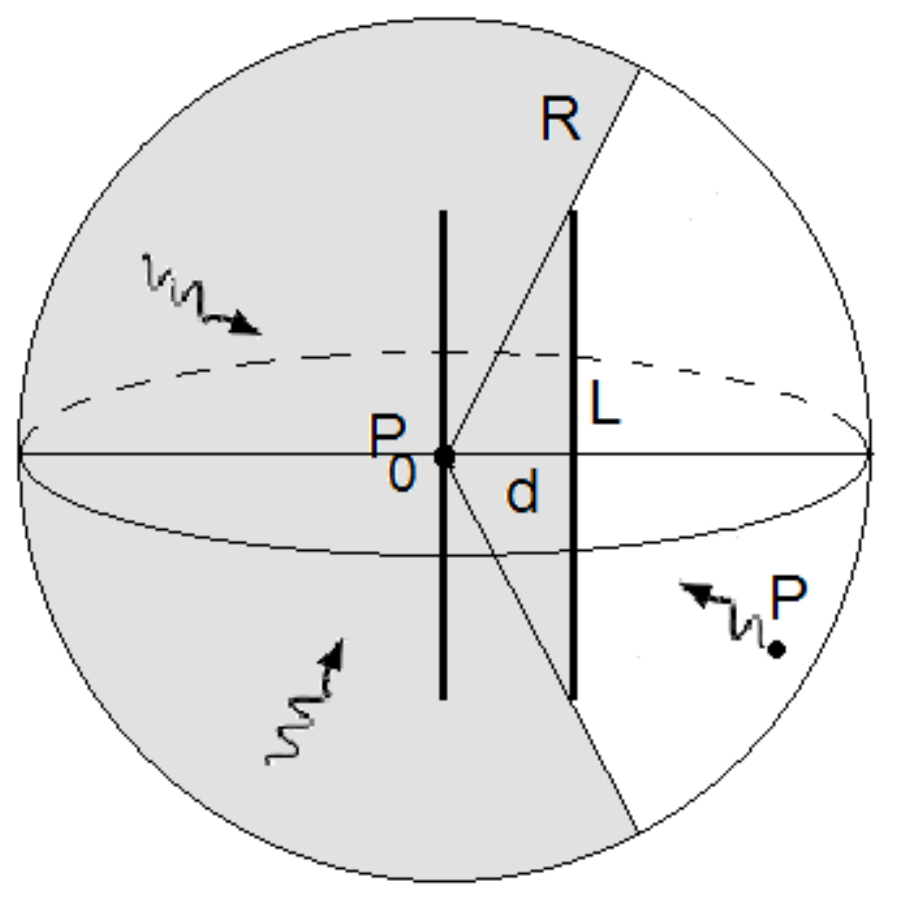}
 \renewcommand{\baselinestretch}{0.8}
 \caption{Heuristic derivation of the zero-point energy for two parallel plates at distance $d$.
 The sphere of radius $R$ represents the whole
 space. The only photons which are allowed to impact on $P_0$ are those in the shadowed volume.}
 \label{figuragine}
 \end{figure}

\subsection{Casimir effect from Heisenberg uncertainty principle}
\label{CEFHP}
In Ref.~\cite{Gine}, the Casimir effect is derived
from the idea that the contribution to the vacuum 
energy at a point $P_0$ of a plate is affected by the presence of the other 
boundary. Specifically, the author considers virtual photons 
produced by vacuum fluctuations somewhere in the space and
arriving at $P_0$. In order to compute the total Casimir energy $\Delta E$,
one has to take into account all the points on the surface $S$ of the plate. 
Therefore, from the HUP 
\begin{equation}
\Delta x\Delta E\simeq \frac{\hslash c}{2}\,, 
\end{equation}
($p=E/c$ for photons), the 
total contribution to the energy fluctuation $\Delta E$ is given
by those photons in a volume $S\Delta x$ around the plate, 
where $\Delta x$ is the position uncertainty of the
single particle. 
Note that, if we had only one plate, $\Delta x$ would be infinite, 
since photons may be created in any point of the space. 
However, this is no longer true in the presence of both
the boundaries. In that case, indeed, 
virtual particles originating from
behind the second plate cannot reach $P_0$. Thus, 
the additional plate acts as a sort of shield.  

The above situation can be depicted as follows:
consider a sphere of radius $R$ 
centered at the point $P_0$ and enclosing both the plates (see Fig.~\ref{figuragine}). 
In the single-plate configuration, the effective volume $S\Delta x$ 
corresponding to the entire space 
can be thought of as the total volume of the sphere
$V_T=4/3\pi R^3$, with $R\rightarrow\infty$. 
Clearly, such a volume will be reduced by including the
second plate, which will prevent particle that pop out in $P$ from
impacting on $P_0$ (with reference to Fig.~2, 
the effective volume is represented by the shadowed region). As a result, we can write 
$S\Delta x=V_T-V_C$,  
where $V_C$ is the volume shielded by the second plate.
In the case of infinite boundaries, or even better when 
$L/(2d)\rightarrow\infty$, one can show that $V_C=2/3\pi R^3$, yielding~\cite{Gine}
\begin{equation}
\label{adx}
S\Delta x\,\simeq\,\frac{2}{3}\pi R^3\,.
\end{equation}
In the above treatment, no length scale
has been considered, hence the volume $S\Delta x$ 
diverges as the radius $R$ increases. 
To cure such a pathological behavior, in Ref.~\cite{Gine} 
the author introduces
a cutoff $r_e$ representing the
effective distance beyond which
photons have a negligible probability
to reach the plate. In this way, Eq.~\eqref{adx}
can be rewritten as
\begin{equation}
\label{adxrenorm}
S\Delta x\,\simeq\,\frac{2}{3}\pi r_e^3\,,
\end{equation}
which has indeed a finite value. Combining this relation with the HUP, 
we then obtain
\begin{equation}
\label{casre}
|\Delta E(r_e)|\,=\,\frac{3}{4}\frac{S\hspace{0.2mm}\hslash\hspace{0.2mm}c}{\pi r_e^3}\,, 
\end{equation}
which implies
\begin{equation}
\label{resimd}
r_e\simeq d\,,
\end{equation}
from comparison
with the exact expression~\eqref{Casenedens} for the Casimir 
energy. Strictly speaking, we would have 
\begin{equation}
\label{setting}
r_e=\frac{\sqrt[3]{540}\hspace{0.2mm}}{\pi}\hspace{0.4mm}d\simeq2.6\,d\,.
\end{equation}

Although the above derivation is straightforward 
and very intuitive, the discussion 
on the physical origin of the length cutoff $r_e$ 
appears to be rather obscure in some points.
Therefore, in order to clarify the meaning of 
Eq.~\eqref{resimd}, let us focus on the
computation of the Casimir 
effect in a simplified
one-dimensional system: similar reasonings
can be readily extended to three dimensions.

\begin{figure}[t]
 \centering
 \includegraphics[width=9.5cm]{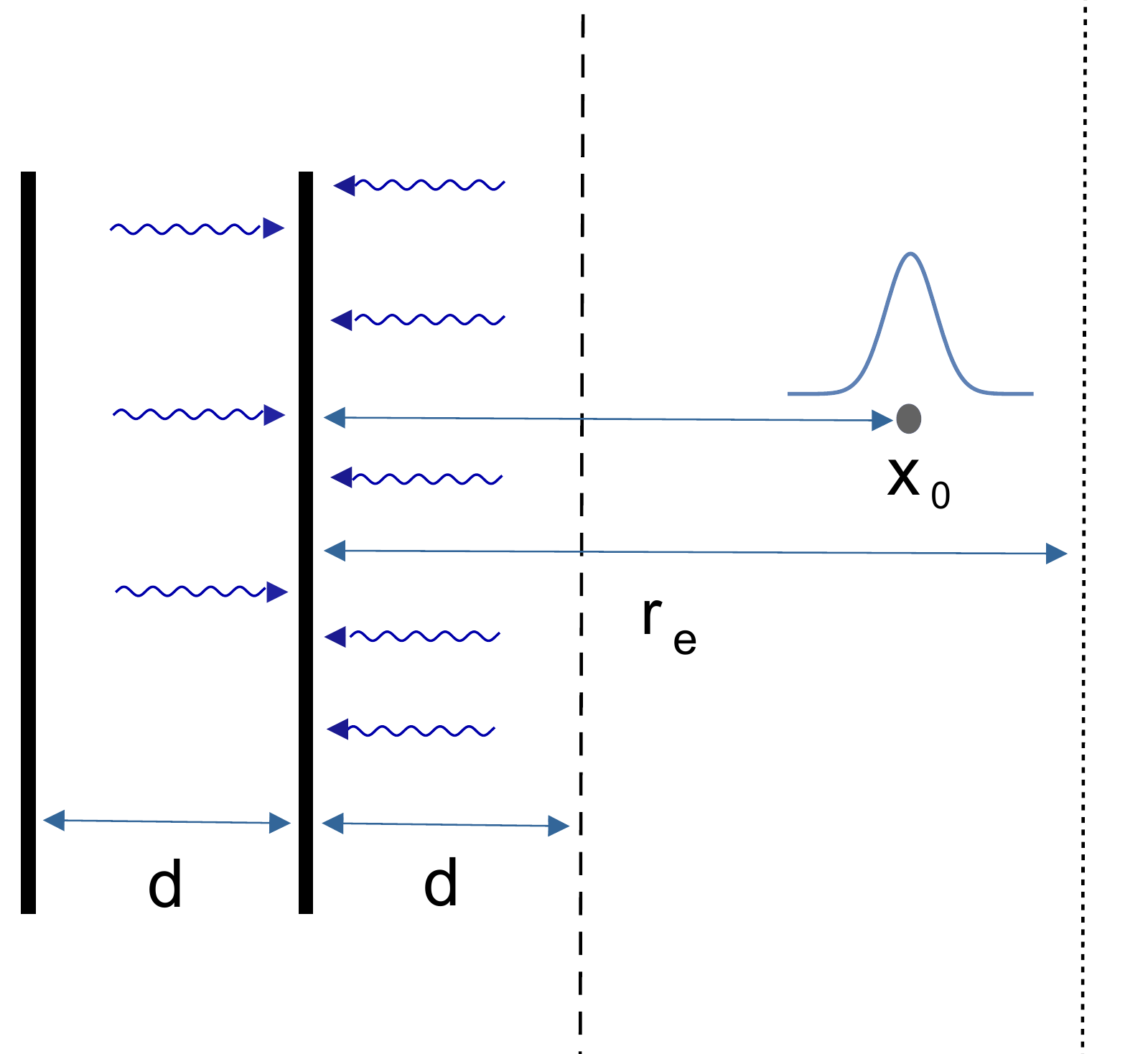}
 \renewcommand{\baselinestretch}{0.8}
 \caption{Setup for the heuristic derivation of the Casimir effect: two infinite parallel plates (bold lines) at distance $d$. The effective radius beyond which the 
creation of virtual photons does not give a significant 
contribution to the Casimir energy is denoted by $r_e$.}
 \label{figura}
 \end{figure}

From the Heisenberg uncertainty relation, 
it is well-known that large energy fluctuations 
live for very short time and, thus, hard 
virtual photons of energy $\Delta E$
can only travel short distances of order
$\hslash c/\Delta E$. As a consequence, the
further these particles are created from a plate, 
the more negligible their contribution to the 
energy around that plate will be. 
Let us apply these considerations 
to the apparatus in Fig.~\ref{figura}.
It is easy to see that virtual photons popping
out in the strip of width $d$ on the right side of the 
right plate do not contribute to the Casimir effect, 
since their pressure is balanced
by those photons originating between the plates.
By contrast, photons coming from a distance
greater than $d$ in the right region do not experience any compensation, 
because their symmetric ``partners'' on the left side
are screened by the first plate. The overall result is a net
force acting on the right plate from right to left.
Of course, this argument can be symmetrically 
applied to the left boundary and provides a qualitative explanation 
for the origin of the attractive Casimir force.

Now, consider a point at a distance 
$x_0>d$ from one of the plates, as in Fig.~\ref{figura}.
Virtual photons originate from quantum
fluctuations in a small region around that point.  
Such a region, however, cannot be smaller than the 
(reduced) Compton length
of the electron, $\lambda_C=\hslash/(m_ec)$, 
otherwise the energy amplitude of the 
fluctuation would exceed 
the threshold $E\simeq m_ec^2$ for the production of 
electron-positron pairs.  
Besides, photons
produced at $x_0$ can impact on the plate
(and therefore contribute to the Casimir force)
only if their energy $E$ is such that $0<E<E_0$, 
where $E_0=\hslash c/x_0$. Particles of higher 
energy $E>E_0$, indeed, would
recombine before reaching the plate, since
the distance they travel is
$x=\hslash c/E<\hslash c/E_0=x_0$. 

We can now assume that photons coming from
$x_0$ originate from fluctuations of energy $E$ with
a probability given by a Boltzmann-like factor
$f(E)=e^{-E/(m_ec^2)}$. Thus, the
total linear energy density (i.e. the energy
per unit length) arriving on the plate will be
\begin{equation}
\label{dens}
|\Delta\varepsilon(E_0)|\,=\,\int_0^{E_0}\frac{dE}{\lambda_C}\frac{E}{m_ec^2}\,f(E)\,=\,\frac{1}{\hslash c}\int_0^{E_0}dE\,E\,e^{-E/(m_ec^2)}\,,
\end{equation}
where, since we are dealing with the electromagnetic
field, we have introduced the natural
threshold of the electron mass/energy $m_ec^2$.
In terms of the distance $x_0$, the above integral 
becomes
\begin{equation}
\label{densbis}
|\Delta\varepsilon(x_0)|\,=\,\hslash c\int_{x_0}^{\infty}\frac{dx}{x^3}\,e^{-\hslash/(m_e\hspace{0.1mm}c\hspace{0.1mm}x)}\,.
\end{equation}
Finally, in order to get the contribution to the Casimir energy
from all the photons which impact on the plate,  
we integrate over all the points $x_0$
such that $d<x_0<\infty$, obtaining 
\begin{equation}
\label{denster}
|\Delta E(d)|\,=\,\int_d^{\infty}dx_0\,\Delta\varepsilon(x_0)\,.
\end{equation}
The integrals in Eqs.~\eqref{densbis} and~\eqref{denster}
can be easily evaluated by observing that, for $x$ large enough, 
the Boltzmann factor $e^{-\hslash/(m_ecx)}$ becomes approximately of
order unity. This yields
\begin{equation}
|\Delta\varepsilon(x_0)|\,\simeq\,\frac{1}{2}\frac{\hslash c}{x_0^2}\,,
\end{equation}
and, hence
\begin{equation}
\label{energiacasimir}
|\Delta E(d)|\,\simeq\,\frac{1}{2}\frac{\hslash c}{d}\,,
\end{equation}
which is in good agreement with the QFT prediction~\eqref{casundim}.
Similarly, one can show that
the generalization to three dimensions leads to a result consistent with 
Eq.~\eqref{Casenedens}.
 
The physical relevance of the above 
discussion becomes clearer if we observe
that probability distributions like those in Eq.~\eqref{dens} 
or~\eqref{densbis} allow us to naturally interpret the effective
radius $r_e$ in Eq.~\eqref{adxrenorm} 
as the distance from the plate below which the vast
majority of photons contribute to the large part of the Casimir energy.
More rigorously, we can define $r_e>d$ as the distance
within which photons carrying the fraction $\gamma$ $(0<\gamma<1)$
of the total Casimir energy are created. 
In other terms, we can write
\begin{equation}
\frac{\hslash c}{2}\int_d^{r_e}\frac{dx}{x^2}\,=\,\gamma\Delta E(d)\,,
\end{equation}
from which
\begin{equation}
\label{25}
r_e\,=\,\frac{d}{1-\gamma}\,.
\end{equation}
Thus, setting $r_e\simeq2.6\hspace{0.5mm}d$ (as
in Eq.~\eqref{setting}) 
amounts to consider
a fraction $\gamma\simeq0.62$
of the total energy responsible for the Casimir effect.

The above picture is quite rough, 
since it relies on the 
adoption of a Boltzmann-like distribution for the energy 
of quantum vacuum fluctuations. As a result, it
underestimates the fraction
of photons produced within the distance $r_e$ from the plate. 
Considerable improvements can be achieved
by employing more realistic functions $f(E)$ in Eq.~\eqref{dens}.
For further details on this topic, see Refs.~\cite{Few}
and therein. 

\subsection{Casimir effect from Generalized uncertainty principle}
Let us now extend the above arguments to the
context of the GUP. As in the previous Subsection, 
we shall focus for simplicity on the one-dimensional case. The 
generalization to three dimensions proceeds in 
a very similar fashion.

We start from the modified uncertainty relation~\eqref{gup}, here recast in the form
\begin{equation}
\label{modforphot}
\Delta x\Delta E\,\simeq\,\frac{\hslash c}{2}\left[1\,+\,\beta{\left(\frac{\Delta E}{E_p}\right)}^2\right].
\end{equation}
By solving with respect to $\Delta E$, we obtain
\begin{equation}
\Delta E\,=\,\frac{\Delta x\hspace{0.2mm}E_p^2}{\hslash\hspace{0.2mm} c\hspace{0.2mm}\beta}\left[1\pm\sqrt{1\,-\,\beta{\left(\frac{\hslash c}{\Delta x\hspace{0.2mm}E_p}\right)}^2}\right],
\end{equation}
where the only solution to be considered is
the one with negative sign, as it reduces to the standard
result for vanishing $\beta$ (conversely, the solution with
positive sign has no evident physical meaning).
After expanding to the first order in $\beta$, it follows that
\begin{equation}
\label{28}
\Delta E\,=\,\frac{\hslash c}{2\Delta x}\left[1+\frac{\beta}{4}{\left(\frac{\hslash c}{\Delta x E_p}\right)}^2\right].
\end{equation}
If we now neglect those photons coming from distances greater
than the effective radius $r_e$, it is natural to assume
the uncertainty position $\Delta x$ of the single photon
to be of the order of $r_e$ and, thus, of $d$, according
to Eq.~\eqref{setting}. Then, by replacing $\Delta x\simeq 2.6\hspace{0.5mm}d$ into Eq.~\eqref{28}, 
the contribution to the Casimir energy at a given point 
reads
\begin{equation}
|\Delta E(d)|\,\simeq\,0.2\hspace{0.4mm}\frac{\hslash c}{d}\left[1\,+\,0.04\hspace{0.3mm}\beta{\left(\frac{\hslash c}{E_p\hspace{0.2mm} d}\right)}^2\right],
\label{betacorrec}
\end{equation}
which indeed agrees with Eq.~\eqref{casundim} 
in the limit $\beta\rightarrow0$.

The above considerations can be now generalized to
three-dimensions by taking into
account the contribution to the zero-point energy at any point of
the plates of area $S$. In doing so, straightforward calculations lead 
to
\begin{equation}
\label{finres}
|\Delta E(d)|\,\simeq\,0.03\hspace{0.4mm}\frac{\hslash\hspace{0.2mm}c\hspace{0.2mm}S}{d^3}\left[1\,+\,0.04\hspace{0.3mm}\beta{\left(\frac{\hslash c}{E_p\hspace{0.2mm}d}\right)}^2\right],
\end{equation} 
which is to be compared with Eq.~\eqref{veinfs}. In 
spite of our skinny assumptions, one can see that 
the obtained expression agrees with the field theoretical result 
as concerns the dependence of the 
GUP correction on the inverse fifth power
of the distance between the plates. 
We also notice that the exact numerical coefficient 
can be recovered by including a proper factor which accounts for
the extension of the GUP~\eqref{modforphot} to
a higher-dimensional system.

\section{Casimir effect for spherical and cylindrical shells: HUP vs GUP approaches}
\label{sphcylshel}
In this Section, we apply our heuristic approach to configurations other than parallel plates. Specifically, we compute the Casimir energy for a spherical and cylindrical shells. It is nevertheless worth mentioning that the following analysis may serve as a basis for the study of more sophisticated geometries too.

\begin{figure}[t]
 \centering
 \includegraphics[width=8.1cm]{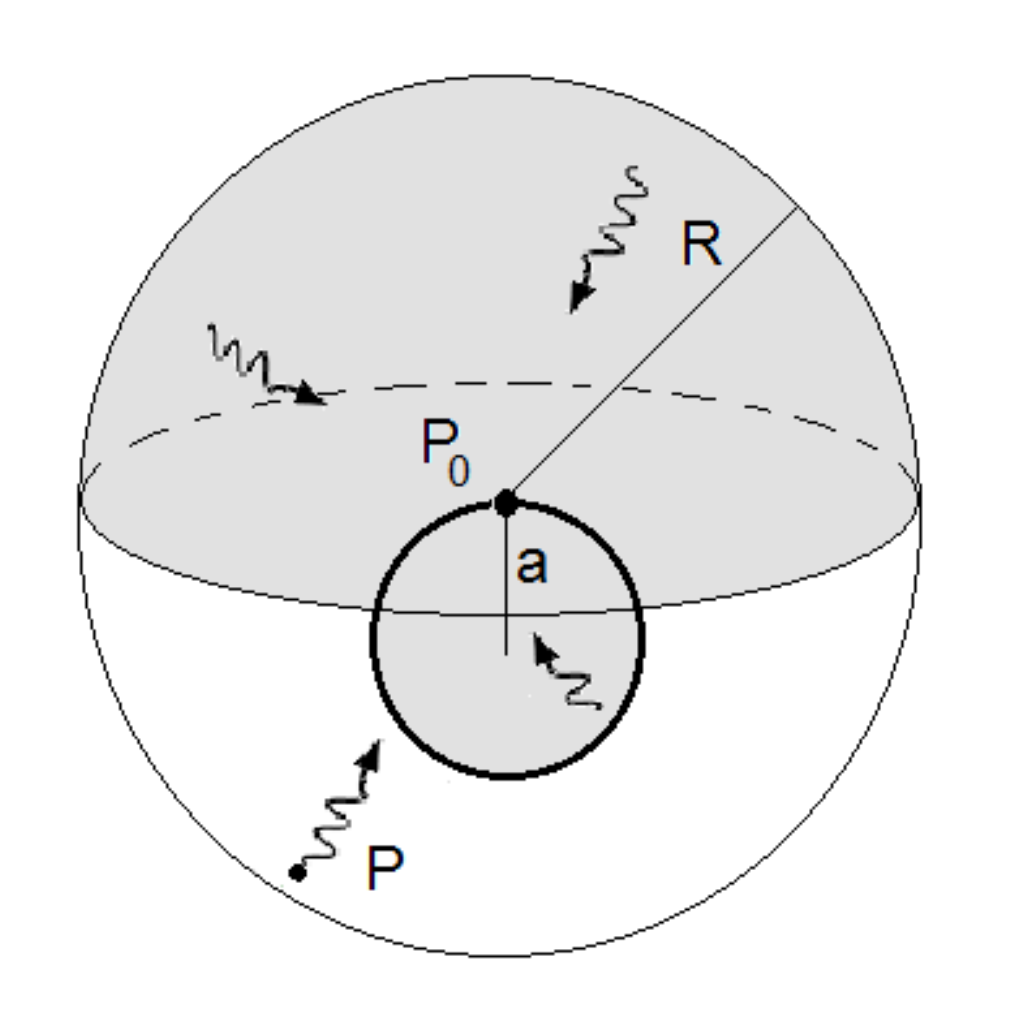}
 \renewcommand{\baselinestretch}{0.8}
 \caption{Heuristic derivation of the zero-point energy for a 
 spherical shell of radius $a$. The outer sphere of radius $R$ represents the whole
 space. The only photons which are allowed to impact on $P_0$ are those in the shadowed volume.}
 \label{sphereb}
 \end{figure}
\subsection{Casimir effect for a spherical shell}
\label{CEFAS}
By following the same reasoning 
as for parallel plates, let us consider a
spherical shell of (finite) radius $a$ enclosed in
a larger sphere of (infinite) radius $R$ representing the whole space (see Fig.~\ref{sphereb}).
By taking a point $P_0$ on the surface of the shell, one can easily understand that 
the only photons which are allowed to impact on $P_0$
are those originating from vacuum fluctuations in the
shadowed volume. 
In light of this, 
we can write the effective volume $S\Delta x$ as the sum of the volume 
of the upper (external) hemisphere and the one of the shell, i.e.
\begin{equation}
\label{effectivevolume}
S\Delta x\,=\,\frac{2}{3}\pi R^3\,+\,\frac{4}{3}\pi a^3\,,
\end{equation}
where $S=4\pi a^2$ is the surface area of the shell. 
For the internal consistency of our formalism, we require $0<a\le R/\sqrt[3]{2}$.
 By drawing a comparison with the parallel-plate system, 
the $a\rightarrow 0$ limit corresponds to the case in which
the two plates are stuck together. In this case, the spherical shell degenerates 
into a single point, since $P_0$ merges with its antipode. 
Consequently, the effective volume $S\Delta x$ will 
be \emph{twice} the volume of the hemisphere of radius $R\rightarrow\infty$, i.e. it will cover all the space\footnote{In order to compute the effective volume
relative to $P_0$ for $a\rightarrow 0$, 
we must also take into account the symmetric contribution of those photons which impact on the antipode of $P_0$.}. 
On the other hand, 
for $a=R/\sqrt[3]{2}\rightarrow\infty$, the point $P_0$
is not affected by the presence of the walls of the shell. This amounts
to the case where the two parallel plates are infinitely far apart from each other. Again, the effective volume will be equal to the whole space, i.e. $S\Delta x=4\pi R^3/3\rightarrow\infty$ from Eq.~\eqref{effectivevolume}.

Now, by using Eq.~\eqref{effectivevolume}, the position uncertainty 
$\Delta x$ reads
\begin{equation}
\label{posuncer}
\Delta x\,=\,\frac{R^3+2a^3}{6\hspace{0.2mm}a^2}\,, 
\end{equation}
that still diverges as $R$ increases. As for the parallel plates, 
however, we can reasonably neglect photons coming from distances
greater than the effective radius $R\sim r_e$. By combining
Eq.~\eqref{posuncer} with the HUP, it follows that
\begin{equation}
|\Delta E (a, r_e)|\,=\,\frac{3\hspace{0.4mm}\hslash c\hspace{0.6mm}a^2}{r_e^3\,+\,2a^3}\,.
\end{equation}
If we now assume $r_e$ to be of the order of the size of the system 
as in Eq.~\eqref{setting}, i.e. $r_e\simeq 2.6\, (2\hspace{0.2mm}a)$, we finally obtain
\begin{equation}
\label{finalsphere}
|\Delta E (a)|\,=\,0.02\,\frac{\hslash c}{a}\,,
\end{equation}
that matches with the QFT result of Refs.~\cite{Boyer,Milton}, up to a factor $1/2$
(as discussed after Eq.~\eqref{25}, one may improve the agreement
by refining the considerations on the photon energy 
distribution as in Eq.~\eqref{dens}). 
Note that we cannot infer any kind of information on the
sign of $\Delta E$, and, thus, on the nature of the Casimir force (whether it is
attractive or repulsive), since our heuristic calculations
only allow to derive the absolute value of the energy shift. 
In this regard, we emphasize that the issue of the sign of the 
zero-point force for a spherical shell is quite controversial. In Refs.~\cite{Boyer,Milton}, 
for instance, it is argued that a conducting sphere 
would tend to be expanded due to effects of vacuum fluctuations. 
On the other hand, if one roughly approximates the sphere 
as two parallel plates of area $S=\pi a^2$ and separation $d\simeq a$ and
considers the standard expression~\eqref{Casenedens} for the energy, 
the opposite sign for the Casimir force would be obtained\footnote{Note that 
the idea to describe a sphere as two (circular) parallel plates 
was originally proposed by Casimir to calculate the fine-structure constant $\alpha$.}~\cite{Casimirsphere}.
A similar result is claimed to be valid for two 
slightly separated hemispheres (that is, a spherical shell sliced
with a very subtle knife) and, more generally, for any symmetric
configuration (see Kenneth-Klick's no-go theorem~\cite{KK}).

\medskip
Let us now investigate to what extent 
the zero-point Casimir energy~\eqref{finalsphere} gets modified by
the Generalized Uncertainty Principle~\eqref{modforphot}. To this aim, 
by replacing Eq.~\eqref{posuncer} into~\eqref{28}, 
we get
\begin{equation}
|\Delta E(a, r_e)|\,=\,\frac{3\hspace{0.3mm}\hslash c\hspace{0.6mm}a^2}{r_e^3\,+\,2a^3}\left\{1\,+\,\beta{\left[\frac{3\hspace{0.3mm}\hslash c\hspace{0.6mm}a^2}{E_p\left(r_e^3\,+\,2a^3\right)}\right]}^2\right\},
\end{equation}
where we have implicitly made use of the 
length cutoff $R\sim r_e$. By setting 
$r_e\simeq 2.6\hspace{0.5mm}(2\hspace{0.2mm}a)$ as before, we find
\begin{equation}
|\Delta E(a)|\,=\,0.02\,\frac{\hslash c}{a}\left[1\,+\,0.0004\hspace{0.2mm}\beta{\left(\frac{\hslash c}{E_p\hspace{0.3mm}a}\right)}^2\right].
\end{equation}
Hence, on the basis of purely heuristic arguments, 
we obtain a GUP term scaling as the inverse cube of the radius of the spherical shell.
As expected, the greater the sphere, the smaller the
correction to the standard (HUP) result. Unlike the parallel-plate configuration, however, 
a full-fledged field theoretical calculation 
has not yet been carried out, thus preventing us
from making any comparison.

\subsection{Casimir effect for a cylindrical shell}
\begin{figure}[t]
 \centering
 \includegraphics[width=8.1cm]{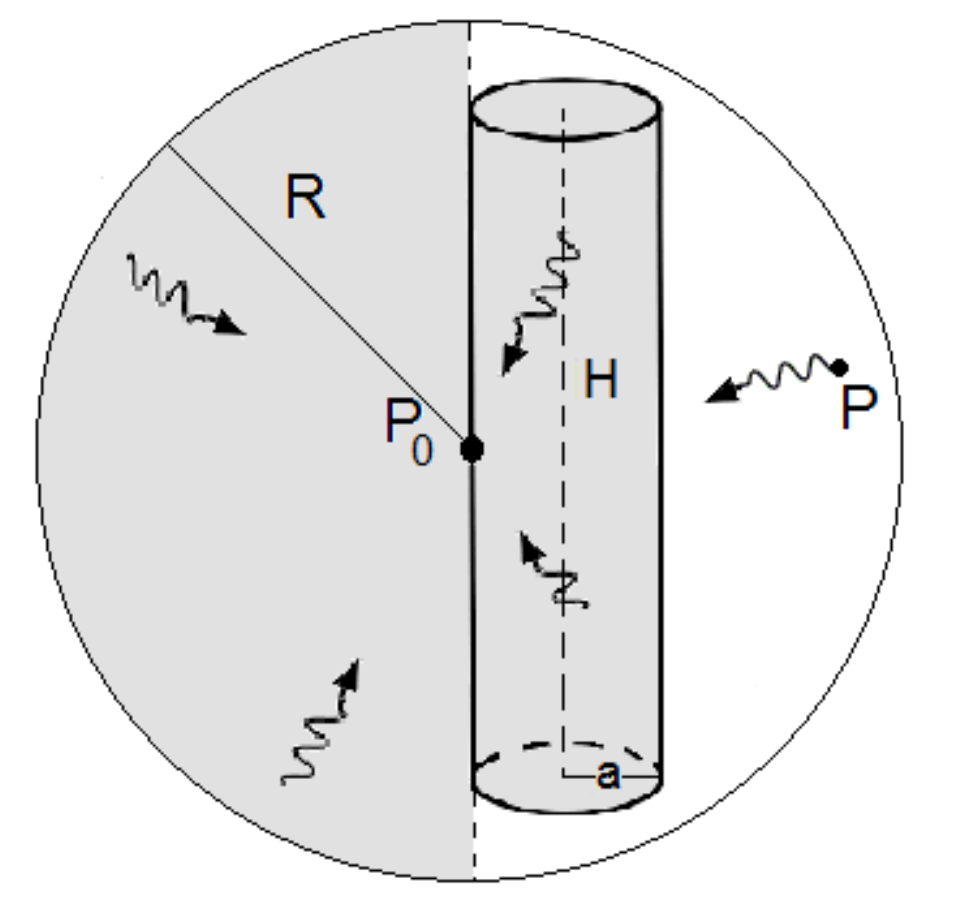}
 \renewcommand{\baselinestretch}{0.8}
 \caption{Heuristic derivation of the zero-point energy for a 
cylindrical shell of radius $a$ and height $H>a$. The sphere of radius $R$ represents the whole space. As for the spherical configuration, the only photons which can reach $P_0$ are those in the shadowed volume.}
 \label{cylinderfig}
 \end{figure}

We now compute the zero-point energy for a cylindrical shell
of radius $a$ and height $H$. As depicted in Fig.~\ref{cylinderfig}, we assume $H>a$; however, 
similar considerations hold true for any size of the system. 

Let us consider a point $P_0$ on the lateral surface
of the cylinder. In this case, we can write the 
effective volume as
\begin{equation}
\label{cyl}
S\Delta x\,=\,\frac{2}{3}\pi {r_e}^3\,+\,\pi a^2\hspace{0mm} H\,,
\end{equation}
where now $S=2\pi\hspace{0.2mm}a\hspace{0.2mm} H$ 
is the lateral surface of the cylinder
and we have already implemented the cutoff $R\sim r_e$
on the radius of the surrounding sphere.
Again, as $a$ and $H$ increase, 
the cylindrical shell tends to cover the whole space, 
while for vanishing $a$ and $H$, it collapses 
in a single point. In both cases, 
the effective volume will be equal to the entire space
(see the discussion after Eq.~\eqref{effectivevolume}).

By inverting Eq.~\eqref{cyl} with respect to the uncertainty
position of photons, we get
\begin{equation}
\label{dxcyl}
\Delta x\,=\,\hspace{0.2mm}\frac{2\hspace{0.2mm}{r_e}^3\,+\,3\hspace{0.2mm}a^2\hspace{0mm} H}{6\hspace{0.2mm}a\hspace{0.2mm} H}\,, 
\end{equation}
which leads to the following expression of the
standard (HUP) energy uncertainty: 
\begin{equation}
|\Delta E (a, H, r_e)|\,=\, \frac{3\hspace{0.4mm}\hslash c\hspace{0.6mm}a\hspace{0.3mm}H}{2\hspace{0.2mm}{r_e}^3\,+\,3\hspace{0.2mm}a^2\hspace{0mm} H}\,.
\end{equation}
Let us now observe that, since we are 
neglecting vacuum fluctuations from distances greater than $r_e$, 
it is reasonable to assume $H\simeq 2\hspace{0.3mm}r_e$
(in other terms, the only photons which can impact on $P_0$ moving 
from the inside to the outside of the cylinder are those in the volume $V\simeq \pi a^2\hspace{0.2mm}(2\hspace{0.2mm}r_e)$). By
setting $r_e\simeq 2.6\, (2\hspace{0.2mm}a)$ as before,  
we then obtain
\begin{equation}
\label{finalcyl}
|\Delta \varepsilon (a)|\,\equiv\,\frac{|\Delta E (a)|}{H}\,=\,0.01\,\frac{\hslash c}{a^2}\,,
\end{equation}
where we have computed the energy per unit length in order to
compare our expression with the QFT result of Ref.~\cite{DeRaad:1981hb}. 
Note that the two outcomes are in good agreement with each other.

\medskip
As usual, corrections induced by the GUP can be estimated
by inserting Eq.~\eqref{dxcyl} into Eq.~\eqref{28}. Straightforward calculations
yield
\begin{equation}
|\Delta \varepsilon(a, r_e)|\,=\,\frac{3\hspace{0.4mm}\hslash c\hspace{0.6mm}a}{{2\hspace{0.3mm}r_e}^3\,+\,6\hspace{0.2mm}a^2\hspace{0.3mm} r_e}\left\{1\,+\,\beta {\left[\frac{6\hspace{0.3mm}\hslash c\hspace{0.6mm}a\hspace{0.6mm}{r_e}}{E_p\left(2\hspace{0.3mm}r_e^3\,+\,6\hspace{0.2mm}a^2\hspace{0.4mm}r_e\right)}\right]}^2\right\},
\end{equation}
which, for $r_e\simeq 2.6\hspace{0.5mm}(2\hspace{0.2mm}a)$, becomes 
\begin{equation}
|\Delta E(a)|\,=\,0.01\,\frac{\hslash c}{a^2}\left[1\,+\,0.01\hspace{0.2mm}\beta{\left(\frac{\hslash c}{E_p\hspace{0.3mm}a}\right)}^2\right].
\end{equation}
As for the sphere, a QFT treatment of Casimir effect
for a cylindrical shell is missing so far.

\section{Discussion and Conclusions}
\label{DandC}
We have computed the corrections to the Casimir energy
in the framework of minimal length theories based
on a generalized uncertainty principle with only a quadratic term in the 
momentum. Calculations have been carried out for three different systems:
the parallel plates, the spherical and cylindrical shells. 
For the first geometry, the result derived via heuristic arguments 
has been compared with the more rigorous
field theoretical expression, showing in both cases  
a dependence of the GUP-correction on the inverse fifth power
of the distance between the plates. On the other hand, 
the absence of a QFT treatment of Casimir effect with GUP
for non-planar configurations does not allow any
comparison for the sphere and the cylinder. Nevertheless, 
such calculations, along with a possible extension of our formalism 
to arbitrary $D$-dimensional systems (see for example Ref.~\cite{Milton}) 
will be investigated in more detail in future works.

Finally, some remarks are in order here. First, we point out that, even though 
direct observations of GUP effects on the
Casimir force are extremely challenging, 
current experiments~\cite{Bressi} might enable us
to fix an upper bound on the minimal
length $(\Delta x)_{\mathrm{min}}=\sqrt{\beta}\hspace{0.2mm}\ell_p$,
and, thus, on the parameter $\beta$.
Furthermore, we emphasize that a similar analysis of
GUP-induced corrections 
has been proposed in Ref.~\cite{SBLC} in the framework
of the Unruh vacuum radiation for accelerated observers.
In that case, the existence of a nonzero minimal length manifests
itself in the form of (in principle) non-thermal corrections to the 
Unruh spectrum, which however can be reinterpreted 
as a shift of the usual Unruh temperature 
for small deformations of the commutator. In passing, we mention
that deviations of the Unruh effect from the
well-known behavior (and, more generally, non-inertial corrections 
on standard predictions of QFT) have been
recently pointed out also in other contexts~\cite{othersect,other2,other3}.
In light of these considerations, it is clear that 
the study of all these unconventional aspects 
of fundamental quantum phenomena represents 
a fertile but still largely uncharted field of research,
since it allows us to test QFT in Planck-scale
regime at both theoretical and experimental levels.
More work is inevitably required along these directions.

\section{Acknowledgements}
The authors would like to thank the anonymous Referee for
his/her comments, which improved the quality of the 
manuscript.

\end{document}